\documentclass[usenatbib,conference]{basi}
\usepackage[british]{babel}
\usepackage[varg]{txfonts}
\usepackage{rotating}
\usepackage{dcolumn}
\usepackage{amsfonts}

\newcommand{\aap}{    {\it Astron. Astrophys.}}
\newcommand{\apj}{    {\it Astrophys. J.}}
\newcommand{\apjl}{   {\it Astrophys. J.}}
\newcommand{\araa}{   {\it Ann. Rev. Astron. Astrophys.}}
\newcommand{\aspcs}{  {\it Astron. Soc. Pac. Conf. Ser.}}
\newcommand{\jpc}{    {\it J. Phys. Conf. Ser.}}
\newcommand{\nat}{    {\it Nature}}
\newcommand{\solphys}{{\it Solar Phys.}}

\begin{document}
\title[Variation in p-modes and Flows with Flaring]{Variations in p-mode Parameters and Sub-surface Flows of Active Regions with Flare Activity}
\author[R.~A.~Maurya and A.~Ambastha]%
       {R.~A.~Maurya\thanks{email: \texttt{ramajor@prl.res.in}} and A.~Ambastha\thanks{email: \texttt{ambastha@prl.res.in}}\\
       Udaipur Solar Observatory, Physical Research Laboratory, Udaipur-313001, India.}

\pubyear{2011}
\volume{00}
\pagerange{\pageref{firstpage}--\pageref{lastpage}}
\conference{Asia-Pacific Solar Physics Meeting}

\date{Received \today}

\maketitle
\label{firstpage}

\begin{abstract}
We examine the characteristic properties of photospheric p-modes and sub-photospheric flows of active regions (ARs) observed during the period of 26-31 October 2003. Using ring diagram analysis of Doppler velocity data obtained from the Global Oscillations Network Group (GONG), we have found that p-mode parameters evolve with ARs and show a strong association with flare activity. Sub-photospheric flows, derived using inversions of p-modes, show strong twist at the locations of ARs, and large variation with flare activity. 
\end{abstract}


\section{Introduction}
\label{S-Intro}

Solar photospheric oscillations having a period of five minutes were first discovered by \citet{Leighton1962}. These arise due to pressure waves (or p-modes) which are trapped in different cavities in the solar interior \citep{Ulrich1970, Leibacher1971, Deubner1975}. The upper boundary of the cavities lies close to the photosphere while the lower one lies at depths depending upon the wavelength of the acoustic wave. The global modes of oscillations have lifetimes long enough to travel completely around the solar circumference and self interfere without suffering a loss of phase coherency greater than $\pi$/2 \citep{Hill1995}. Accurate frequency measurement of these modes provides the global properties of solar interior, such as temperature, pressure, chemical compositions, etc. But they are insensitive to local characteristics, viz., meridional and complex flows of ARs. On the other hand, high degree ($\ell>$300) p-modes are more sensitive to local properties of ARs. Therefore, photospheric and sub-photospheric properties of ARs are studied using local oscillations by local helioseismology techniques \citep{Lindsey1993}.

Local helioseismology provides a three-dimensional view of ARs. Three main techniques are used: ring diagram analysis \citep{Hill1988}, time-distance method \citep{Duvall1993} and acoustic holography \citep{Lindsey1990}. The first one is generalization of the global helioseismology over small area of ARs compared to the whole Sun. This is based on the reasonably well understood physics of normal modes, while the interpretation of other techniques are still in developement \citep[see][and references therein]{Gizon2010}. These methods have been extensively used earlier to study the photospheric p-modes and sub-photospheric flows of ARs \citep[see][and references therein]{Maurya2010}.  

Solar p-modes are believed to be caused by stochastic excitation in the convection zone but they are expected to be modified by energetic transients, such as, flares and CMEs. \citet{Wolff1972a} first suggested that a large flare can modify the p-mode amplitude by exerting mechanical impulse of the thermal expansion on the photosphere. The amplification is expected to be larger in high degree p-modes, because they are concentrated near the photosphere, and their typical wavelengths approximately matches the scale of the pulse. Subsequently, flare related variations in high degree p-mode parameters have been observationally reported by several researchers \citep{Ambastha2003a, Ambastha2004a, Howe2004, Maurya2009a}.

ARs are the most important candidates for energetic solar transients. Therefore, it is interesting to study their internal structure and dynamics. Local helioseismology provides a unique observational tool to determine  sub-photospheric flows of ARs by inverting the photospheric p-modes \citep[see][and references therein]{Maurya2010}. Helioseismic studies show that sunspots are rather shallow, near-photospheric phenomena \citep{Kosovichev2000, Basu2004} and are locations of large scale flows at the photosphere \citep{Haber2002, Braun2004, Zhao2004a, Komm2005b, Maurya2010e, Maurya2010a, Maurya2011}.

Aim of this study is two fold: First we try to understand the characteristics of p-mode parameters at locations of active and quiet regions, and their variations with flaring activity. Secondly, we examine sub-photospheric flow variations in ARs from non-flaring to flaring phases. We have carried out these studies for the active and quiet regions that appeared during 26 - 31 October 2003. The paper is organized as follow: In Section~\ref{S-ARFlare}, we describe the ARs observed during the aforesaid period. Section~\ref{S-DataAna} discusses the observational data and methods of analysis. Section~\ref{S-ResDisc} presents results of our analysis. Finally, summary and conclusions are provided in  Section~\ref{S-SumConcl}.            

\section{Active Regions and Flares}
\label{S-ARFlare}

\begin{figure}\centering
\includegraphics[width=1.0\textwidth,clip=,bb=11 3 461 209]{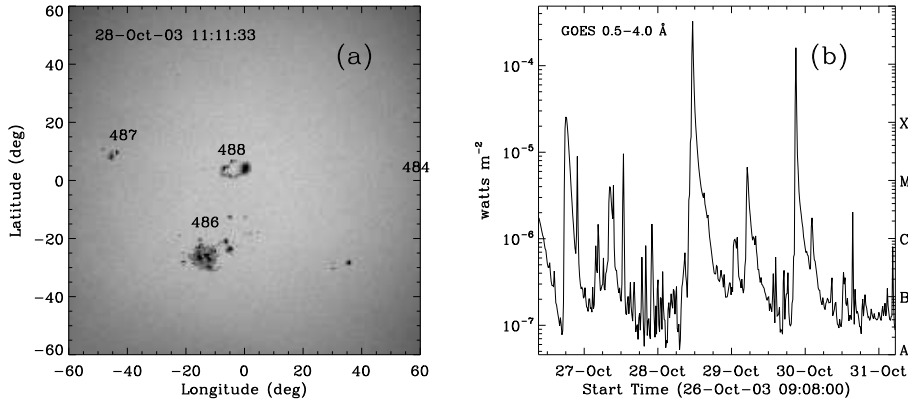}
\caption{(a) MDI continuum image of the Sun observed at 11:11:13\,UT/28 October 2003. NOAA ARs are labeled on their appropriate locations. (b) Integrated GOES X-ray light curve in the wavelength range 0.5-4.0\AA. \label{F-ContImgGOES}}
\end{figure}

During the period of 26-31 October 2003, several ARs appeared on the solar disk, but the ARs NOAA 10484, 10486, and 10488 dominated due to their complex and large size. Amonst these, NOAA 10486 was the largest and most flare productive as it gave rise to flares of largest magnitude, viz., X17/4B, X10/2B, of the solar cycle 23. These ARs are shown in a MDI continuum image observed at 11:11:33\,UT/28 October 2003 (Figure~\ref{F-ContImgGOES}(a)). The transient activities for the aforesaid time period are shown by the integrated GOES X-ray light curve in the wavelength range 0.5--4.0\AA~ in Figure~\ref{F-ContImgGOES}(b).
   
\section{Observational Data and Analysis}
\label{S-DataAna}

Observational data for the study of ARs for the period 26-31 October 2003 were obtained from the GONG archive. The photospheric p-mode parameters were computed using ring diagram analysis \citep{Hill1988} of the data cube ($16^{\rm o}\times16^{\rm o}\times1664^{\rm m}$). Then, we computed horizontal components ($u_{\rm x}$, $u_{\rm y}$) of sub-photospheric flows as a function of depth from inversions of p-modes \citep{Gough1985}. The vertical component ($u_{\rm z}$) of flow was calculated using continuity equation \citep{Komm2004}. Maps of kinetic helicity density (KHD), a measure of twist of sub-photospheric flows, were then constructed using these three components of flows. 
 
\section{Results and Discussions}
\label{S-ResDisc}

The results obtained from the analysis of ARs for the period 26-31 October 2003 are shown in Figure~\ref{F-ContImgGOES}-~\ref{F-MKhMap}.

\subsection{Variations in p-mode Parameters of Active Regions}
\label{Sb-pmode}

\begin{figure}
\centering
\includegraphics[width=1.0\textwidth,clip=,bb=52 26 497 218]{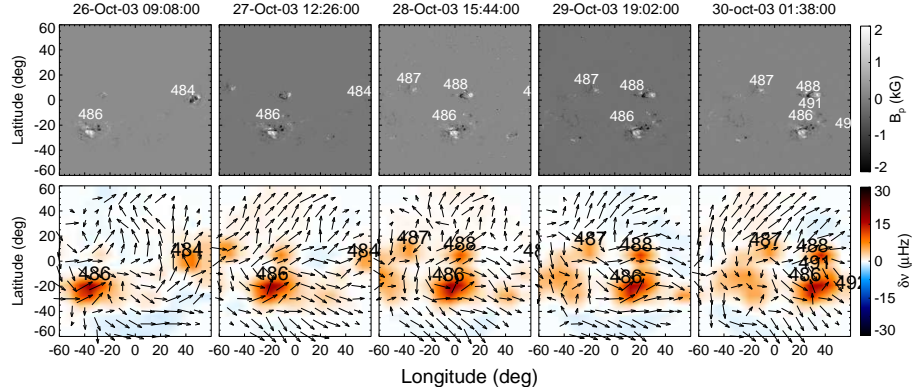}
\caption{MDI magnetograms (\textit{top}) and averaged p-mode (n = 0 - 5) parameters (\textit{bottom}) in the longitude and latitude range $\pm60^{\rm o}$ for the period 26-31 October 2003. In the bottom, background image corresponds to the frequency shift and vectors represent the horizontal flows.\label{F-PModeMap}}
\end{figure}

Figure~\ref{F-PModeMap}(\textit{top}) shows the MDI magnetograms in the range $\pm60^{\rm o}$ of longitude and latitude during the period 26-31 October 2003. The bottom panel shows the co-aligned maps of the p-mode parameters: frequency shift~ $\delta\nu$ in the background half-tone image, and horizontal components of flows Ux and Uy marked by vectors. The ARs are labeled in both panels at their appropriate places. The start date/time of the ring-diagram data cubes corresponding to the maps are given in the top of the figure. It would be interesting to examine the p-mode characteristics as follows: (i) compare the flare productive ARs with dormant and quiet regions, and (ii) examine their spatio-temporal evolution. For this, we have computed the mode parameters averaged over radial orders n = 0--5 in the range $\pm60^{\rm o}$ of longitudes and latitudes.

In Figure~\ref{F-PModeMap}(\textit{bottom}), vectors represent the horizontal flows and background red (blue) colors show the positive (negative) frequency shift in the averaged modes. It is evident that there were large, positive frequency shifts at the AR's locations, which evolved as the ARs evolved. For instance, NOAA 10488 was initially associated with small positive $\delta\nu$ value on 26 October 2003 which increased till 30 October 2003 as the AR grew in complexity and size. Interestingly, the AR was intensely flare productive during this period.

\begin{figure}
\centering
\includegraphics[width=1.0\textwidth,clip=,bb=33 22 505 140]{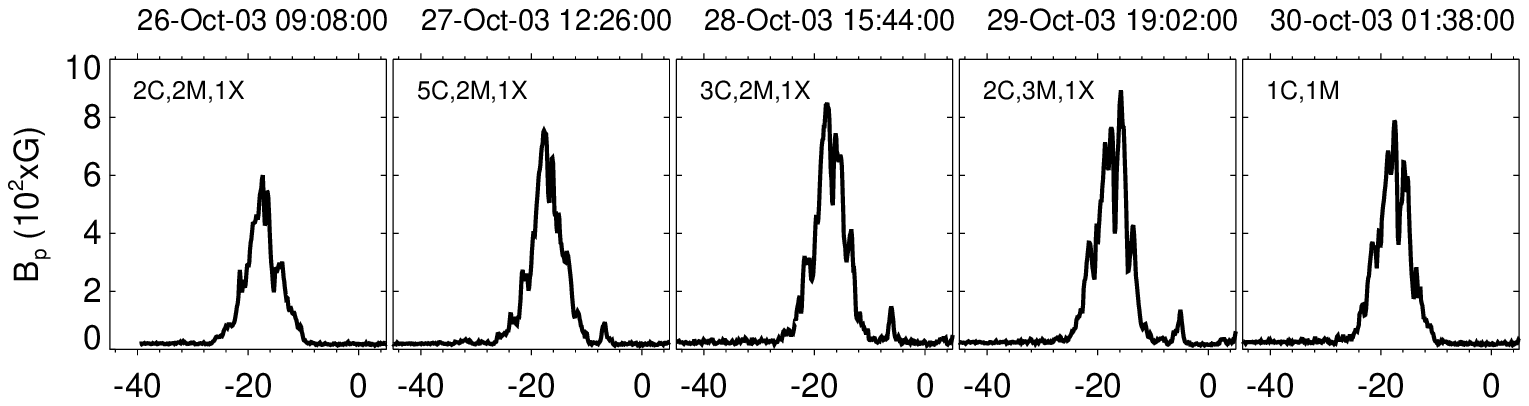}\\
\includegraphics[width=1.0\textwidth,clip=,bb=30 29 505 235]{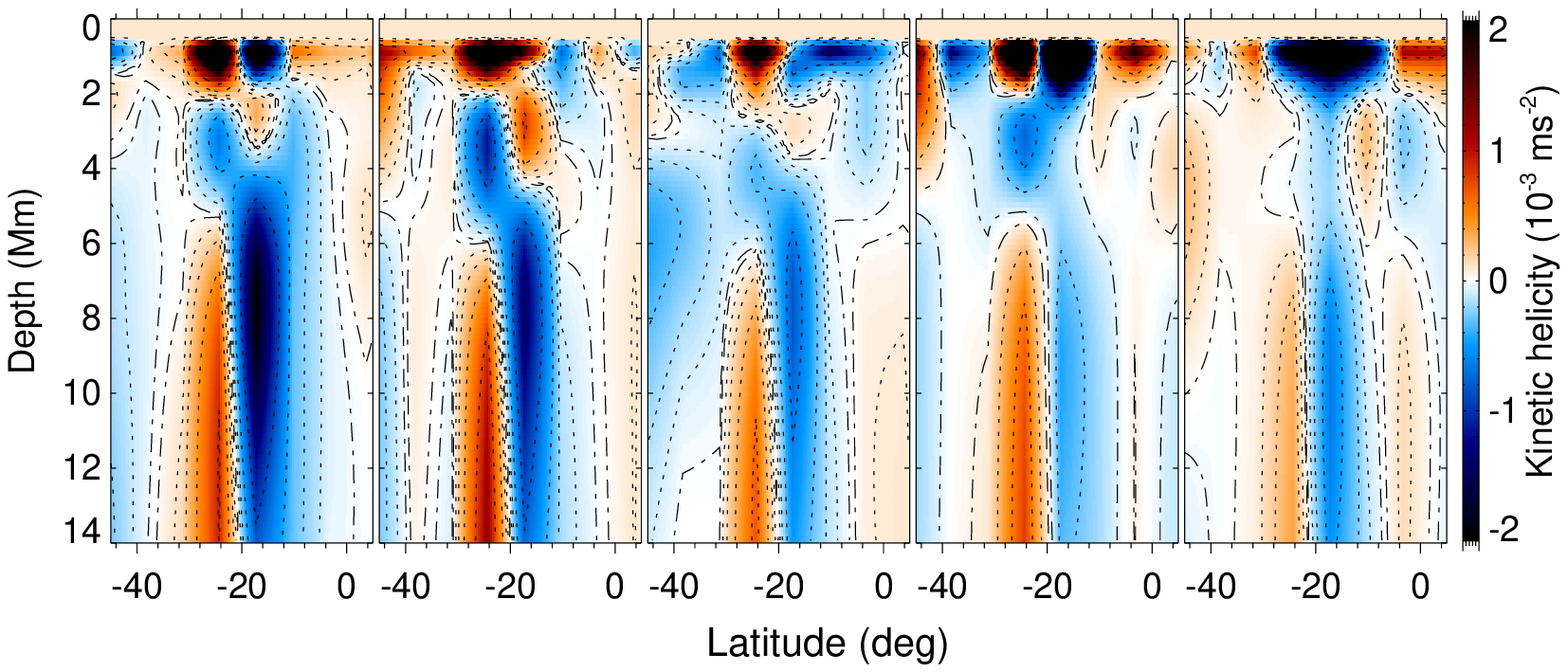}
\caption{Evolution of (\textit{top}) magnetic flux  and (\textit{bottom}) kinetic heliciy density  during 26-31 October 2003.\label{F-MKhMap}}
\end{figure}

It is evident that horizontal flows showed a large deviation from the general flow pattern around the site of AR NOAA 10486 as seen from the changing direction of vectors. Since Ux and Uy are the weighted averages over depth, the flow pattern reveals highly sheared flows in the interior of NOAA 10486 as compared to other less flare active ARs and the surroundings (see Figure~\ref{F-MKhMap} \textit{top}). These results provide further confirmation on the association of flare productivity of NOAA 10486 with p-mode parameters, sub-photospheric flows and their variations.   

\subsection{Variations in Sub-Photospheric Flows of Active Regions}
\label{Sb-flow}

Energetic transients are caused by the changes in magnetic fields which are rooted beneath the photosphere. The sub-photospheric flows result in braiding and intertwining of the rising magnetic flux tubes. Therefore, we expect the changes in magnetic field configuration to be governed by sub-photospheric flows. In order to investigate this issue, we have determined kinetic helicity of sub-photospheric flows corresponding to the images shown in Figure~\ref{F-PModeMap}.

Kinetic helicity density maps, as a function of depth and latitude for a fixed Carrington longitude 285$^{\rm o}$, are shown in Figure~\ref{F-MKhMap}(\textit{bottom}) during 26-31 October 2003. Contour levels are drawn at the 0.5, 2.5, 5, 10, 20, 40, 60, 80$\%$ of the absolute maximum of the maps. Top panel shows magnetic flux as a function of latitude corresponding to the \textit{bottom panel}. The number of flares in C, M and X class of GOES SXR are given in the \textit{top panel}. The meridian for Carrington longitude 285$^{\rm o}$ passes through the center of NOAA 10486 (see Figure~\ref{F-PModeMap}). Therefore, large magnetic flux (\textit{top}) and kinetic helicity density (\textit{bottom}) represents the location of NOAA 10486. 

Figure~\ref{F-MKhMap} shows a large twisted flow beneath the ARs as compared to the quiet regions. The locations of large KHD (\textit{bottom}) coincide well with the large magnetic flux areas. Magnetic flux in AR NOAA 10486 increased till 29 October 2003 and decayed thereafter. This AR was extremely flare productive during 28 and 29 October 2003 and produced several helioseismic signatures corresponding to its major flares \citep{Donea2005, Kosovichev2006, Maurya2009a}. On 26 October 2003, it possessed a large KHD, increasing on 27-28 October 2003, when this AR produced the X17/4B flare. After showing a decaying trend, it increased again during 29 October 2003, when it produced another large X10/2B flare. Thereafter, it simplified during 30-31 October 2003. Such systematic variation in internal flows reveals a strong role of fluid/magnetic topology contributing in the initiation of large flares. 

\section{Summary and Conclusions}
\label{S-SumConcl}

From the study of p-modes and sub-photospheric flows beneath ARs that appeared on the solar disk during 26-31 October 2003, we have found the following important results: i) Large p-mode frequency shifts occurred at locations of ARs as compared to the quiet regions, revealing presence of large surface and sub-surface flows in ARs. ii) Sub-surface flows of ARs were comparatively more complex and twisted. iii) Large systematic variations in internal flow  parameters were found associated with flaring activity.

\label{lastpage}
\end{document}